# Dual Time-Space Model of Wave Propagation

Alexei Krouglov

*Matrox Graphics Inc.*

*3500 Steeles Ave. East, Suite 1300, Markham, Ontario L3R 2Z1, Canada*

Email: Alexei.Krouglov@matrox.com

# ABSTRACT

Here to represent the propagation of waves I attempted to describe them separately in space and time domains. The time and space wave equations are obtained and investigated, and formulas for the wave propagation are expressed. I also tried to apply the model to the description of such physical phenomena as the constancy of the speed of light and propagation of ocean waves – tsunami.

*Keywords*: Wave Equation; the Speed of Light; Tsunami

## 1. Model Assumptions

For generality, I will consider here 'the energy's waves'. It gives me the possibility to abstract from some insignificant details (from the model's point of view).

For simplicity, I will describe the one-dimensional space. One can look at the coordinate as a vector to get a picture in multi-dimensional space.

I denote $U(x,t)$ to be the value of energy in point $x$ at time $t$. I also denote $\Phi(x,t)$ to be the level of energy in point $x$ at time $t$ (coincides with $U(x,t)$ when the system is at rest).

Difference between the value $U(x,t)$ and the level $\Phi(x,t)$ constitutes the energy's disturbance $\Delta U(x,t)$, i.e.

$$\Delta U(x,t) = U(x,t) - \Phi(x,t). \qquad (1)$$

When system is at rest, we can say

$$\Delta U(x,t) \equiv 0 \qquad (2)$$

or equivalently

$$U(x,t) \equiv \Phi(x,t). \qquad (3)$$

I don't mean in (3) that system has to be static to reside at rest. We will consider a moving system in the eighth section of paper.

In contrast to the conventional wave equations, which are described by partial differential equations of second order (see [3, 4, 6, 7]), I will be using the following assumptions in the attempt to describe the dynamics of energy's disturbance separately in time and in space.

In the time domain,

(1) The second derivative of energy's value with respect to time is inversely proportional to energy's disturbance.

(2) The first derivative of energy's level with respect to time is directly proportional to energy's disturbance.

In the space domain,

(3) The second derivative of energy's value with respect to direction is inversely proportional to energy's disturbance.

(4) The first derivative of energy's level with respect to direction is directly proportional to energy's disturbance.

To simplify the reasoning I also introduce the dual variables $P(x,t)$, which I will describe in the next two sections.

## 2. Model Description

Until time $t_0 = 0$ I assume that system was at rest, i.e.

$$U(x,t) = \Phi(x,t) = \Phi_0 \qquad (4)$$

for $-\infty < t < 0$ and for all $x$.

At time $t_0 = 0$ the system was excited in the point $x_0 = 0$.

Thus we can write

$$U(0,0) = \Phi_0 + \Delta\Phi_0. \qquad (5)$$

Let me try to describe the dynamics of disturbance

$$\Delta U(x,t) = U(x,t) - \Phi(x,t),$$

for $0 \leq t < +\infty$, $0 \leq x < +\infty$, $\Delta U(0,0) = \Delta \Phi_0$, $\dfrac{d\Delta U(0,0)}{dt} = 0$, and $\dfrac{d\Delta U(0,0)}{dx} = 0$.

In the time domain,

$$\frac{dP(x_1,t)}{dt} = -\lambda_1 (U(x_1,t) - \Phi(x_1,t)), \tag{6}$$

$$\frac{d^2 U(x_1,t)}{dt^2} = \lambda_2 \frac{dP(x_1,t)}{dt}, \tag{7}$$

$$\frac{d\Phi(x_1,t)}{dt} = -\lambda_3 \frac{dP(x_1,t)}{dt}, \tag{8}$$

where $x_1 \geq 0$, $\lambda_1, \lambda_2, \lambda_3 > 0$ are constants, and $P(x_1,t)$ is dual variable, which connects the second derivative of $U(x_1,t)$ and first derivative of $\Phi(x_1,t)$ with disturbance $\Delta U(x_1,t)$.

We can solve equations (6) – (8) (see [1, 5]) to obtain the dynamics of disturbance $\Delta U(x_1,t)$ in the time domain,

$$\frac{d^2}{dt^2} \Delta U(x_1,t) + \lambda_1 \lambda_3 \frac{d}{dt} \Delta U(x_1,t) + \lambda_1 \lambda_2 \Delta U(x_1,t) = 0. \tag{9}$$

(a) If $\dfrac{\lambda_3}{2} > \sqrt{\dfrac{\lambda_2}{\lambda_1}}$ the solution of (9) is

$$\Delta U(x_1,t) = C_1 e^{k_1 t} + C_2 e^{k_2 t}, \tag{10}$$

where $C_1$ and $C_2$ are constants of integration, and

$$k_{1,2} = \lambda_1 \left( -\frac{\lambda_3}{2} \pm \sqrt{\left(\frac{\lambda_3}{2}\right)^2 - \frac{\lambda_2}{\lambda_1}} \right).$$

Since both $k_1 < 0$ and $k_2 < 0$, the value $\Delta U(x_1,t) \to 0$ for $t \to +\infty$, and oscillations don't take place in the time domain.

(b) If $\dfrac{\lambda_3}{2} = \sqrt{\dfrac{\lambda_2}{\lambda_1}}$ the solution of (9) is

$$\Delta U(x_1,t) = (C_1 + C_2 t) e^{-\frac{\lambda_1 \lambda_3}{2} t}. \qquad (11)$$

Again the value $\Delta U(x_1,t) \to 0$ for $t \to +\infty$, and oscillations are not here in the time domain.

(c) If $\dfrac{\lambda_3}{2} < \sqrt{\dfrac{\lambda_2}{\lambda_1}}$ the solution of (9) is

$$\Delta U(x_1,t) = e^{-\frac{\lambda_1 \lambda_3}{2} t} (C_1 \cos \beta t + C_2 \sin \beta t), \qquad (12)$$

where $\beta = \lambda_1 \sqrt{\dfrac{\lambda_2}{\lambda_1} - \left(\dfrac{\lambda_3}{2}\right)^2}$.

Here the value $\Delta U(x_1,t) \to 0$ for $t \to +\infty$, and we have the situation of damped oscillations in the time domain.

Now let me do the same for energy's disturbance in the space domain,

$$\frac{dP(x,t_1)}{dx} = -\mu_1 (U(x,t_1) - \Phi(x,t_1)), \qquad (13)$$

$$\frac{d^2 U(x,t_1)}{dx^2} = \mu_2 \frac{dP(x,t_1)}{dx}, \qquad (14)$$

$$\frac{d\Phi(x,t_1)}{dx} = -\mu_3 \frac{dP(x,t_1)}{dx}, \qquad (15)$$

where $t_1 \geq 0$, $\mu_1, \mu_2, \mu_3 > 0$ are constants.

Then dynamics of disturbance $\Delta U(x,t_1)$ in the space domain is,

$$\frac{d^2}{dx^2}\Delta U(x,t_1)+\mu_1\mu_3\frac{d}{dx}\Delta U(x,t_1)+\mu_1\mu_2\Delta U(x,t_1)=0. \tag{16}$$

(d) If $\frac{\mu_3}{2} > \sqrt{\frac{\mu_2}{\mu_1}}$ the solution of (16) is

$$\Delta U(x,t_1) = C_1 e^{l_1 t} + C_2 e^{l_2 t}, \tag{17}$$

where $l_{1,2} = \mu_1\left(-\frac{\mu_3}{2} \pm \sqrt{\left(\frac{\mu_3}{2}\right)^2 - \frac{\mu_2}{\mu_1}}\right)$.

Since both $l_1 < 0$ and $l_2 < 0$, the value $\Delta U(x,t_1) \to 0$ for $x \to +\infty$, and oscillations don't take place in the space domain.

(e) If $\frac{\mu_3}{2} = \sqrt{\frac{\mu_2}{\mu_1}}$ the solution of (16) is

$$\Delta U(x,t_1) = (C_1 + C_2 x) e^{-\frac{\mu_1\mu_3}{2}x}. \tag{18}$$

Again the value $\Delta U(x,t_1) \to 0$ for $x \to +\infty$, and oscillations are not here in the space domain.

(f) If $\frac{\mu_3}{2} < \sqrt{\frac{\mu_2}{\mu_1}}$ the solution of (16) is

$$\Delta U(x,t_1) = e^{-\frac{\mu_1\mu_3}{2}x}(C_1 \cos\gamma x + C_2 \sin\gamma x), \tag{19}$$

where $\gamma = \mu_1\sqrt{\frac{\mu_2}{\mu_1} - \left(\frac{\mu_3}{2}\right)^2}$.

Here the value $\Delta U(x,t_1) \to 0$ for $x \to +\infty$, and we have the situation of damped oscillations in the space domain.

Earlier I said about intention to represent the dynamics of disturbance $\Delta U(x,t)$ from initial values $\Delta U(0,0) = \Delta \Phi_0$, $\frac{d\Delta U(0,0)}{dt} = 0$, and $\frac{d\Delta U(0,0)}{dx} = 0$.

Let me show how the value $\Delta U(x_1, t_1)$ is obtained for $x_1 > 0$ and $t_1 > 0$.

If I denote $c$ the velocity of disturbance traveling, then disturbance will reach the point $x_1$ at time $t_0 = x_1/c$.

Hence we have $\Delta U(x_1, t_1) \equiv 0$ for $t_1 \leq t_0$.

For $t_1 = t_0 + \Delta t$ (where $\Delta t > 0$) we may find at first $\Delta U(0, \Delta t)$ from equation (9), and initial values $\Delta U(0,0) = \Delta \Phi_0$ and $\frac{d\Delta U(0,0)}{dt} = 0$. Then we find the value $\Delta U(x_1, \Delta t)$ from equation (16), and initial values $\Delta U(0, \Delta t)$ and $\frac{d\Delta U(0, \Delta t)}{dx} = 0$.

Thus the obtained value $\Delta U(x_1, \Delta t)$ is the energy's disturbance in the point $x_1$ at the time $\Delta t$ for ideal *'instant'* velocity of disturbance traveling that is equal to the value of energy's disturbance in the point $x_1$ at the time $(t_0 + \Delta t)$ for disturbance traveling with the real velocity $c$.

Note we used above the boundary condition $\frac{d\Delta U(0,t)}{dx} \equiv \frac{d\Delta U(0,0)}{dx} = 0$, which is easy to understand through the energy *'invariant'* described in the fourth section.

## 3. Dual Model

Here we describe dynamics of dual variable $P(x,t)$.

At first in the time domain let me differentiate (6) and integrate (7). It gives us,

$$\frac{d^2 P(x_1,t)}{dt^2} + \lambda_1\lambda_3 \frac{dP(x_1,t)}{dt} + \lambda_1\lambda_2 P(x_1,t) + C_3 = 0, \qquad (20)$$

where $C_3 = \lambda_1 \left( \frac{dU(x_1,t_0)}{dt} - \lambda_2 P(x_1,t_0) \right)$. Note that the values $\frac{dU(x_1,t_0)}{dt}$ and $P(x_1,t_0)$ are taken at time $t_0$ when energy's disturbance reaches the point $x_1$.

We can simplify (20) by changing the variable,

$$P_1(x_1,t) = P(x_1,t) + \frac{1}{\lambda_2} \cdot \frac{dU(x_1,t_0)}{dt} - P(x_1,t_0).$$

It will give us,

$$\frac{d^2 P_1(x_1,t)}{dt^2} + \lambda_1\lambda_3 \frac{dP_1(x_1,t)}{dt} + \lambda_1\lambda_2 P_1(x_1,t) = 0. \qquad (21)$$

Equation (21) can be solved similarly to (9). Hence we obtain $P_1(x_1,t) \to 0$ for $t \to +\infty$, and therefore $P(x_1,t) \to P(x_1,t_0) - \frac{1}{\lambda_2} \cdot \frac{dU(x_1,t_0)}{dt}$ when $t \to +\infty$ for all $x_1 > 0$.

Similarly in the space domain,

$$\frac{d^2 P(x,t_1)}{dx^2} + \mu_1\mu_3 \frac{dP(x,t_1)}{dx} + \mu_1\mu_2 P(x,t_1) + C_4 = 0, \qquad (22)$$

where $C_4 = \mu_1 \left( \dfrac{dU(x_0, t_1)}{dx} - \mu_2 P(x_0, t_1) \right)$. Note that the values $\dfrac{dU(x_0, t_1)}{dx}$ and $P(x_0, t_1)$ are taken either (a) at the time $t_1$ in the excited point $x_0$ for energy's disturbance traveling with ideal 'instant' velocity, or (b) at the time $\left( t_1 - \dfrac{x - x_0}{c} \right)$ in the excited point $x_0$ for energy's disturbance traveling with real velocity $c$.

Therefore $P(x, t_1) \to P(x_0, t_1) - \dfrac{1}{\mu_2} \cdot \dfrac{dU(x_0, t_1)}{dx}$ when $x \to +\infty$ for all $t_1 > 0$.

## 4. Energy Invariant

We can see that oscillations involve transformations of energy's value into energy's 'velocity' in the time domain and energy's 'tangent' in the space domain.

Therefore we can deduce the following energy's 'invariant'.

In the time domain,

$$\varepsilon(x_1, t) = \sqrt{(\Delta U(x_1, t))^2 + \left( \dfrac{1}{\beta} \cdot \dfrac{d}{dt} \Delta U(x_1, t) \right)^2}, \qquad (23)$$

and in the space domain,

$$\varepsilon(x, t_1) = \sqrt{(\Delta U(x, t_1))^2 + \left( \dfrac{1}{\gamma} \cdot \dfrac{d}{dx} \Delta U(x, t_1) \right)^2}. \qquad (24)$$

Here I mean that if 'damping' coefficients $\lambda_3$ and $\mu_3$ were zeroes then both values $\varepsilon(x_1, t)$ and $\varepsilon(x, t_1)$ would be constants.

Then in both time and space domains energy 'invariant' would look like,

$$\varepsilon(x,t) = \sqrt{(\Delta U(x,t))^2 + \left(\frac{1}{\beta} \cdot \frac{d}{dt} \Delta U(x,t)\right)^2 + \left(\frac{1}{\gamma} \cdot \frac{d}{dx} \Delta U(x,t)\right)^2}. \tag{25}$$

In other words, without *'damping'* coefficients we would have infinite oscillations both in time and in space.

## 5. Interference of Energy's Waves

I will consider the situation where the system was initially at rest, and later it was excited twice.

First time it was done in point $x_0 = 0$ at time $t_0 = 0$, where $\Delta U(0,0) = \Delta \Phi_0$, $\dfrac{d\Delta U(0,0)}{dt} = 0$, and $\dfrac{d\Delta U(0,0)}{dx} = 0$. Second time it was in point $x_0' = \rho > 0$ at time $t_0' = \tau > 0$. I assume that system was excited second time before the energy's first disturbance reaches the point $x_0'$, therefore $\tau < \rho/c$, $\Delta U(\rho, \tau) = \Delta \Phi_0'$, $\dfrac{d\Delta U(\rho, \tau)}{dt} = 0$, and $\dfrac{d\Delta U(\rho, \tau)}{dx} = 0$.

Let me track the value $\Delta U(x_1, t_1)$ for $x_1 > \rho$ and $t_1 > \tau$.

We have $\Delta U(x_1, t_1) \equiv 0$ if $t_1 \leq \tau + \dfrac{x_1 - \rho}{c}$.

For $\left(\tau + \dfrac{x_1 - \rho}{c}\right) < t_1 \leq \dfrac{x_1}{c}$, we will find the value $\Delta U(x_1, t_1)$ from initial values $\Delta U(\rho, \tau)$, $\dfrac{d\Delta U(\rho, \tau)}{dt}$, and $\dfrac{d\Delta U(\rho, \tau)}{dx}$ similarly as it was done in the second section. We only have to apply here different time interval,

$$\Delta t = t_1 - \tau - \frac{x_1 - \rho}{c}.$$

For $t_1 > \frac{x_1}{c}$, the energy's disturbance from both points $x_0 = 0$ and $x'_0 = \rho$ will reach the point $x_1$. At first, we may obtain the value $\Delta U_1(x_1, t_1)$ from initial values $\Delta U(\rho, \tau)$, $\frac{d\Delta U(\rho, \tau)}{dt}$, and $\frac{d\Delta U(\rho, \tau)}{dx}$ (using the time interval

$\Delta t_1 = t_1 - \tau - \frac{x_1 - \rho}{c}$). Then we can find the value $\Delta U_2(x_1, t_1)$ from initial values

$\Delta U(0,0)$, $\frac{d\Delta U(0,0)}{dt}$, and $\frac{d\Delta U(0,0)}{dx}$ (where the time interval is $\Delta t_2 = t_1 - \frac{x_1}{c}$).

Finally we calculate the united value $\Delta U(x_1, t_1)$ by adding both partial values together, i.e. $\Delta U(x_1, t_1) = \Delta U_1(x_1, t_1) + \Delta U_2(x_1, t_1)$.

## 6.  Generalization of 'Boundary' Conditions

We have described so far the situation when system was excited in some point $x_0$ with non-zero energy's disturbance $\Delta U(x_0, t_0) \neq 0$ and zero disturbance rate $\frac{d\Delta U(x_0, t_0)}{dt} = 0$.

Assuming the non-zero disturbance rate in initial point $x_0 = 0$ and time $t_0 = 0$ doesn't bring big changes in our previous reasoning. We have to consider additionally the initial value $\frac{d\Delta U(x_0,t_0)}{dt} \neq 0$ when we are solving equation (9) in the second section.

Let me look now at case when the system is excited in some interval $[0,a]$ at initial time $t_0 = 0$. We may assume the regular condition with $\Delta U(0,0) = \Delta \Phi_0$, $\frac{d\Delta U(0,0)}{dt} \neq 0$, $\frac{d\Delta U(0,0)}{dx} \neq 0$ and $\Delta U(a,0) = 0$, $\frac{d\Delta U(a,0)}{dt} \neq 0$, and $\frac{d\Delta U(a,0)}{dx} = 0$.

Other points in the interval $(0,a)$ have initial values somewhere in the middle.

When we calculate the value $\Delta U(x_1,t_1)$ for $x_1 > a$ and $t_1 > 0$, the interference of energy's waves has to be taken into account.

Hence, for $t_1 \leq \frac{x_1 - a}{c}$ we have,

$$\Delta U(x_1,t_1) \equiv 0. \tag{26}$$

Interference of energy's waves appears for bigger $t_1$. To simplify its representation I denote $\Delta \Psi(x,t,\rho,\tau)$ the value of energy's disturbance $\Delta U(x,t)$, which spreads to the point $x$ and time $t$ from initial values $\Delta U(\rho,\tau)$, $\frac{d\Delta U(\rho,\tau)}{dt}$, and $\frac{d\Delta U(\rho,\tau)}{dx}$.

Then for $\frac{x_1 - a}{c} < t_1 \leq \frac{x_1}{c}$ we can write,

$$\Delta U(x_1,t_1) = \int_{x_1 - ct_1}^{a} \Delta \Psi(x_1,t_1,\xi,0)d\xi. \tag{27}$$

For $t_1 > \frac{x_1}{c}$ it takes place,

$$\Delta U(x_1,t_1) = \int_0^a \Delta\Psi(x_1,t_1,\xi,0)d\xi. \tag{28}$$

## 7. Disturbance Traveling and Wave Propagation

We have to distinct the concepts of velocity $c$ of disturbance's traveling and velocity of wave propagation.

The former is an external value for our model while the latter one is determined inside the model.

Indeed the velocity $w$ of wave propagation is the ratio of the period of oscillations in space, and the period of oscillations in time (or inverse ratio of the corresponding frequencies).

Thus $w = \dfrac{T_{space}}{T_{time}} = \dfrac{\beta}{\gamma}$.

Hence

$$w = \frac{\lambda_1}{\mu_1}\sqrt{\frac{\dfrac{\lambda_2}{\lambda_1} - \left(\dfrac{\lambda_3}{2}\right)^2}{\dfrac{\mu_2}{\mu_1} - \left(\dfrac{\mu_3}{2}\right)^2}}, \tag{29}$$

and when *'damping'* coefficients $\lambda_3 = 0$ and $\mu_3 = 0$, we have,

$$w = \sqrt{\frac{\lambda_1 \lambda_2}{\mu_1 \mu_2}}. \tag{30}$$

## 8.  Example One: Invariance of the Speed of Light

I want to underline again that velocity $w$ of propagation of energy's disturbance is not the same as velocity $c$ of traveling of energy itself.

This point becomes important when we address the phenomenon of the speed of light.

If me move the system associated with energy's level $\Phi(x,t)$ with non-zero velocity then the velocity $w$ of wave propagation will remain the same in new system, and will be described by equation (29).

Here I see the reason of negative outcome of Michelson-Morley experiment, which started the creation of the theory of relativity [2, 3, 4].

## 9.  Example Two: Tsunami or 'Harbor Wave'

It looks possible to apply the model to describe the phenomenon of *'tsunami'*.

An earthquake passes its energy to ocean water. This energy is transformed into elevating of water column.

Using equations (17) – (19) we can obtain the value $\Delta U(x,t_1)$ of energy's disturbance traveling on significant distances from the earthquake's place.

Then we can calculate the height, on which the water column of particular depth has to be elevated to convey the required amount of energy.

That value gives us the height of tsunami.

**References**

1. V. I. Arnol'd, "Ordinary differential equations," 3$^{rd}$ edition, Springer Verlag, Berlin; New York, 1992.

2. A. Einstein et al., "The Principle of Relativity," Dover, New York, 1952.

3. W. Elmore and M. Heald, "Physics of Waves," McGrow – Hill, New York, 1969.

4. D. Menzel (ed.), "Fundamental Formulas of Physics," Vols. I and II, Dover, New York, 1960.

5. N.S. Piskunov, "Differential and Integral Calculus," Groningen P. Noordhoff, 1965.

6. S.L. Sobolev, "Partial Differential Equations of Mathematical Physics," Pergamon Press, Oxford, UK, 1964.

7. A.N. Tikhonov and A.A. Samarskii, "Partial Differential Equations of Mathematical Physics," Holden – Day, San Francisco, 1964.